\documentstyle[aps,pre,eqsecnum,epsf,graphicx]{revtex}
\begin{document}

\draft

\title{Tracking the phase-transition energy in disassembly of hot nuclei}

\author{C. B. Das$^1$, S. Das Gupta$^1$, L. Beaulieu$^2$,
T. Lefort$^3$ \footnote{Present address: Laboratoire de Physique
Corpusculaire de Caen, F-14050 Caen Cedex, France},
K. Kwiatkowski$^3$ \footnote{Present address: Los Alamos National
Laboratory, Los Alamos, NM 87545}, V. E. Viola$^3$,
S. J. Yennello$^4$, L. Pienkowski$^5$, R. G. Korteling$^6$,
and H. Breuer$^7$}

\address{$^1$
Physics Department, McGill University, Montr{\'e}al, Canada H3A 2T8,}
\address{$^2$Department de Physique, Universit{\'e} Laval, Quebec, 
Canada G1K 7P4}
\address{$^3$Department of Chemistry and IUCF, Indiana University, Bloomington,
Indiana 47405}
\address{$^4$Department of Chemistry and Cyclotron Laboratory, Texas A $\&$ M
University, College Station, Texas 77843}
\address{$^5$Heavy Ion Laboratory, Warsaw University, 02-093 Warsaw, Poland}
\address{$^6$Department of Chemistry, Simon Fraser University, Burnaby, British
Columbia, Canada V5A 1S6}
\address{$^7$Department of Physics, University of Maryland, College Park,
Maryland 20742}

\date{\today}

\maketitle

\begin{abstract}
In efforts to determine phase transitions in 
the disintegration of highly excited heavy nuclei,
a popular practice is to parametrise the yields
of isotopes as a function of temperature
in the form $Y(z)=z^{-\tau}f(z^{\sigma}(T-T_0))$, where $Y(z)$'s 
are the measured yields and $\tau, \sigma$ and $T_0$ are fitted to
the yields. Here $T_0$ would be interpreted
as the phase transition temperature. For finite systems such as those obtained
in nuclear collisions, this parametrisation is only approximate
and hence allows for extraction of $T_0$ in more than one way. In this work
we look in detail at how values of $T_0$ differ, depending on methods
of extraction. It should be mentioned that
for finite systems, this approximate parametrisation works not only at 
the critical point, but also for first order phase transitions (at least
in some models). Thus the approximate fit is no guarantee that one is
seeing a critical phenomenon.

A different but more conventional search for the nuclear phase transition
would look for a maximum in the specific heat
as a function of temperature $T_2$. 
In this case $T_2$ is interpreted
as the phase transition temperature.  Ideally $T_0$ and $T_2$
would coincide.  We invesigate this possibility, both in theory and from 
the ISiS data, performing both canonical ($T$) and microcanonical ($e=E^*/A$)
calculations.

Although more than one value of $T_0$ can be extracted from the approximate
parmetrisation, the work here points to the best value from among the
choices. Several interesting results, seen in theoretical calculations,
are borne out in experiment.

\end{abstract}


\section{Introduction}

In studies of phase transitions in the disintegration of highly excited 
heavy nuclei, a popular path for deducing the occurrence of a phase transition
is to examine the yields of composites.  These are readily available 
from experimental data and hence have been the focus of many theoretical
studies \cite{Dasgupta1}.  The usual practice is \cite{Stauffer} to
use a parametrisation
\begin{eqnarray}
Y(z)=z^{-\tau}f(z^{\sigma}(T-T_0))
\end{eqnarray}
and extract values of $\tau,\sigma$ and $T_0$, which occur
in models of critical phenomena \cite{Stauffer}. The parameters
$\tau$ and $\sigma$
are critical exponents and $T_0$ is the critical temperature.
Alternately, in a microcanonical
formalism one would write
\begin{eqnarray}
Y(z)=z^{-\tau}f(z^{\sigma}(e-e_0))
\end{eqnarray}
Here $e=E^*/A$, the excitation energy per nucleon, and $e_0$
would be the phase transition energy.  Formulae (1.1) and 
(1.2) assume that the thermodynamic limit is reached.  In practice,
in the nuclear case we have a finite system that disintegrates 
and thus the above parametrisation is only approximate.  Hence the values
of the parameters can be extracted in more than one way and these values
may not be the same. We point out that it is largely the Coulomb force 
that leads different methods to give two different values for 
$e_0$, labeled as $e_1$ and $e_1'$.

Alternative but perhaps more common
tools for studies of phase transition in other fields of physics
are measurements of compressibility, specific heat etc.  
Experimental data for specific heat were studied
in the nuclear case and was indeed the cause of great 
excitement \cite{Poch}.

We have therefore two distinct ways of trying to deduce a phase
transition energy: from the distribution of composites  
as the excitation energy is varied (as explained, even here there 
can be more than one value)
or, what may be more difficult but achievable, to locate an extremum
of the specific heat.  We label the excitation energy at which the specific
heat maximises as $e_2$.
It is not obvious that the values of $e_1,e_1'$ and $e_2$ are close,
although from the seminal work of Coniglio-Klein
\cite {Co} on clusterisation this result could be anticipated.
   
We have compared both approaches in the nuclear case in the
framework of two
models.  Although the models are very different and each has its own
strengths and weaknesses, both reveal the following interesting 
features.  If we switch off the Coulomb interaction between protons,
the deduced phase transition energies, $e_1, e_1'$ and $e_2$ are close.
With the inclusion of the Coulomb force, $e_1$ and $e_1'$ begin
to diverge.  For a nucleus of the size of $^{197}$Au, the case we
study and for which fragmentation data exist, the difference in
the $e_1$ and $e_1'$ is significant.  Further, one of these values stays
close to the value at which the specific heat
maximises and gives a good measure of the phase transition energy.

The two models we use are the Lattice Gas Model (LGM) \cite
{Dasgupta1,Pan}
and a thermodynamic model \cite {Dasgupta2,Bhatt}.  The second model
is close in spirit to the Statistical Multifragmentation Model
of Copenhagen \cite {Bondorf}.  We
choose to use a microcanonical simulation for LGM.  So here the primary
quantity is the excitation energy $e$ per particle and a temperature
can be derived afterwards \cite{Das1}. 
For the thermodynamic model,
we do a canonical calculation so that the temperature is the primary
parameter and an excitation energy $e$ per particle can be derived
afterwards.

In section 2 we give details of the LGM calculations.  Results of the 
thermodynamic 
model are presented in section 3.  In section 4 we investigate the ISiS 
data within those formalisms.
We present summary and conclusion in section 5.

\section{Results from LGM}
Numerical techniques for microcanonical simulations with LGM have been
published
\cite{Das1}.  Calculations are done for fixed $E=Ae$ where $e$ is the
excitation energy per nucleon.  This is the primary quantity for 
simulations.  The temperature for each simulation can be calculated
from $T=<2E_{kin}>/3$.  This is discussed in detail in \cite{Das1}.  
For more discussions about the LGM with Coulomb force we
refer to \cite{Sam}, section II.  Bonds due to nuclear forces
are taken to be -5.33 MeV between unlike particles and 0 between
like particles.  The LGM has several drawbacks,
the most noticeable being the lack of quantum effects, which leads
to an incorrect caloric curve near $T=0$.  The LGM has the following advantages
not shared by several other models.  It includes interactions
between composites.  It incorporates the Coulomb
interaction in a much more basic fashion (at the nucleonic level)
than several other models.
This is very important for us since this work points to a new effect
brought about solely by the Coulomb interaction. 
Also the LGM produces particle-stable composites
\cite{Sam,Campi} so that the complicated problem of subsequent
particle evaporation is circumvented.

All calculations reported here are for $Z=79$ and $N=118$ ($^{197}$Au).
At each total energy we compute averages after 50,000 simulated events.
We use $9^3$ ($\rho/\rho_0=.27$) lattice sites. 

The extraction of parameters from yields (Eq.(1.2)) merits consideration.
Discussions of this parametrisation can be found in \cite{Stauffer} where
it is used to model a continuous phase transition in an infinite
system.  As already stated,
one does not expect the above parametrisation to be exact except 
in the thermodynamic limit.  For very finite systems as is the case with
disintegrating nuclei formed in very energetic nuclear collisions, 
the parametrisation
is only approximate and is by no means a signature of a critical
phenomenon but rather that of a phase transition in a finite system,
first order or otherwise \cite{Das2}.
Theory demands that $z$ of the yields be
not small.  In the nuclear case it is also not too big since the
disassembling system itself is very finite.  We limit $z$ between 3
and 17, which is similar to most published work on the subject.
Since the fit is expected to be only approximate,
there is more than one prescription for getting the ``best''
parameters.  Here we follow the prescription given 
in \cite{Gulminelli}:

(1) If Eq. (1.2) were exact, then at $e=e_0$ we would have
$\sum (ln Y(z)-ln C+\tau ln z)^2=0$, as each individual term 
in the sum would be zero.  Of course Eq.(1.2) is not exact and
thus the sum above will not be zero at any value of $e$.  However,
this is a valid question to ask.  At any given $e$ how well does
the distribution fit a power law and what is the value of $\tau$
that gives a best fit to a power law ? At each $e$ we
get a ``best'' $\tau$ by a least-square fit, i.e., by minimising
$\sum (ln Y(z)-ln C +\tau lnz)^2$ with respect to $\tau$ and $C$.
The ``goodness'' of fit is given by the smallness of the sum
which we define to be $\chi^2$ (there are other definitions of
$\chi^2$)
\begin{eqnarray}
\chi^2\equiv \frac{1}{N}\sum(ln Y(z)-lnC+lnz)^2
\end{eqnarray}
Here $N$ is the number of terms in the sum.  
From this step we have a ``best'' $\tau$ and a $\chi^2$ {\it vs.} $e$.
One obvious choice of $e_0$ is that value of $e$ where $\chi^2$ is
minimum (see \cite{Scharenberg}).  The value of $e_0$ deduced using this
criterion will be called $e_1'$.  While this is quite reasonable, it does not
use the scaling property $z^{\sigma}(e-e_0)$ at this stage at all.
That is left upon an optimum choice of $\sigma$ afterwards.  The more
complicated procedure that is followed below is designed to give better
scaling properties.

The ``best'' $\tau$ {\it vs.} $e$ curve will usually have a minimum which
we call $\tau_{min}$.

(2) Define $q=z^{\sigma}(e-e_0)$; $f(q)$
has a maximum for some value of $q=\tilde{q}$: $f_{max}=f(\tilde{q})$.
For each $z$ the yield $Y(z)$ as a function of $e$ has a maximum at
some value of $e_{max}(z)$.  At this excitation energy $Y(z)_{max}=
z^{-\tau}f_{max}$ where $f_{max}$ is a constant independent of $z$.
This allows us to choose values for $\tau$ and $f_{max}$ using a
$\chi^2$ test.

(3) The value of $\tau$ found above will be higher than $\tau_{min}$.  This
means if we look for $e$ appropriate for $\tau$, two values are
available from the $\tau$ vs. $e$ curve (see Fig. 1).  
The lower value is chosen as 
the value of $e_0$.  The scaling property is badly violated by the
other choice.  The value of $e_0$ chosen by this prescription will
be labelled by $e_1$.

(4) Now that we know $e_0=e_1$ and $e_{max}(z)$, the excitation at which
each $z$ is maximised, we find form a least squares fit the
value of $\sigma$ from the condition $z^{\sigma}(e_{max}-e_1)=const$.
for all $z$.

(5) The scaling law can now be tested by plotting $Y(z)z^{\tau}$ {\it vs.}
$z^{\sigma}(e-e_1)$.  Plots for all $z$ should fall on the same graph.

\vskip 0.2in
\epsfxsize=3.5in
\epsfysize=5.0in
\centerline{\rotatebox{270}{\epsffile{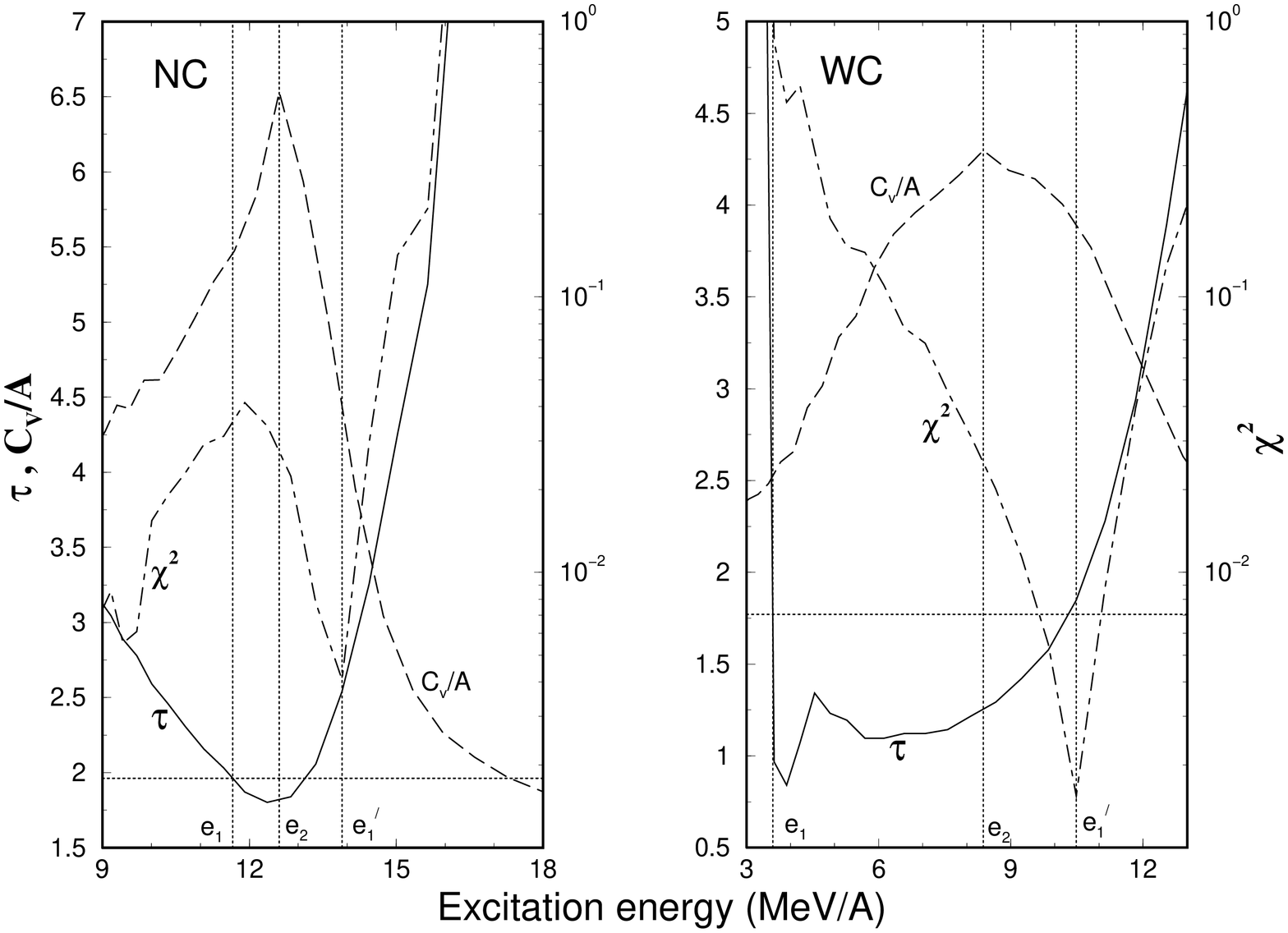}}}
\vskip 0.4in

Fig.1: {\it In this figure we show ``best'' $\tau$ and $\chi^2$ {\it vs.} $e$
(step (1), section II) and also $C_V$ {\it vs.} $e$ for the LGM 
calculation.  The horizontal
line below 2 is a straight line drawn at the value of $\tau$ from step (2),
section II.  The locations $e_1, e_2 $ and $e_1'$ are the values of $e$
for the scaling fit, maximum of specific heat and the minimum of $\chi^2$.
The left panel is an LGM calculation without any Coulomb force, the
right panel is with the inclusion of Coulomb.  Notice the increase of
$\Delta e=e_2-e_1$ when Coulomb is included.}\\

Fig.1 depicts graphs (steps (1) to (3)) with and without the inclusion
of the Coulomb force.  In the graphs we also plot the specific heat
per nucleon.  We refer to the location of the maximum of specific
heat as $e_2$.
The lessons from LGM that we like to emphasize can be learned from Figs.
(1) and (2).  In Fig. 1 consider first the no-Coulomb
case.  Here $e_1$ is 11.66 MeV, $e_2$ is 12.61 MeV, and $e_1'$
(defined by the minimum of $\chi^2$, the prescription of 
\cite{Scharenberg}, see part (1) above) is 13.88 MeV.  We regard them
as close. With Coulomb $e_1$ drops well below $e_2$ and $e_1'$;
$e_2$ and $e_1'$ stay close.

Of the three energies $e_1$, $e_2$ and $e_1'$, which one marks
phase transition better? Without Coulomb, the $Y(z)$ curves at the
three $e$-values are quite similar,
but with Coulomb, $e_1$ is
clearly in the phase co-existence region and is below the
phase transition energy (please see Fig. 2).  
Looking at yields at $e_2$ and $e_1'$, there is no obvious choice  
in deciding which marks the phase transition point better.
However, since an extremum in the
value of specific heat is a standard signature of phase transition,
our preference is with $e_2$. 

\vskip 0.2in
\epsfxsize=3.0in
\epsfysize=4.5in
\centerline{\rotatebox{270}{\epsffile{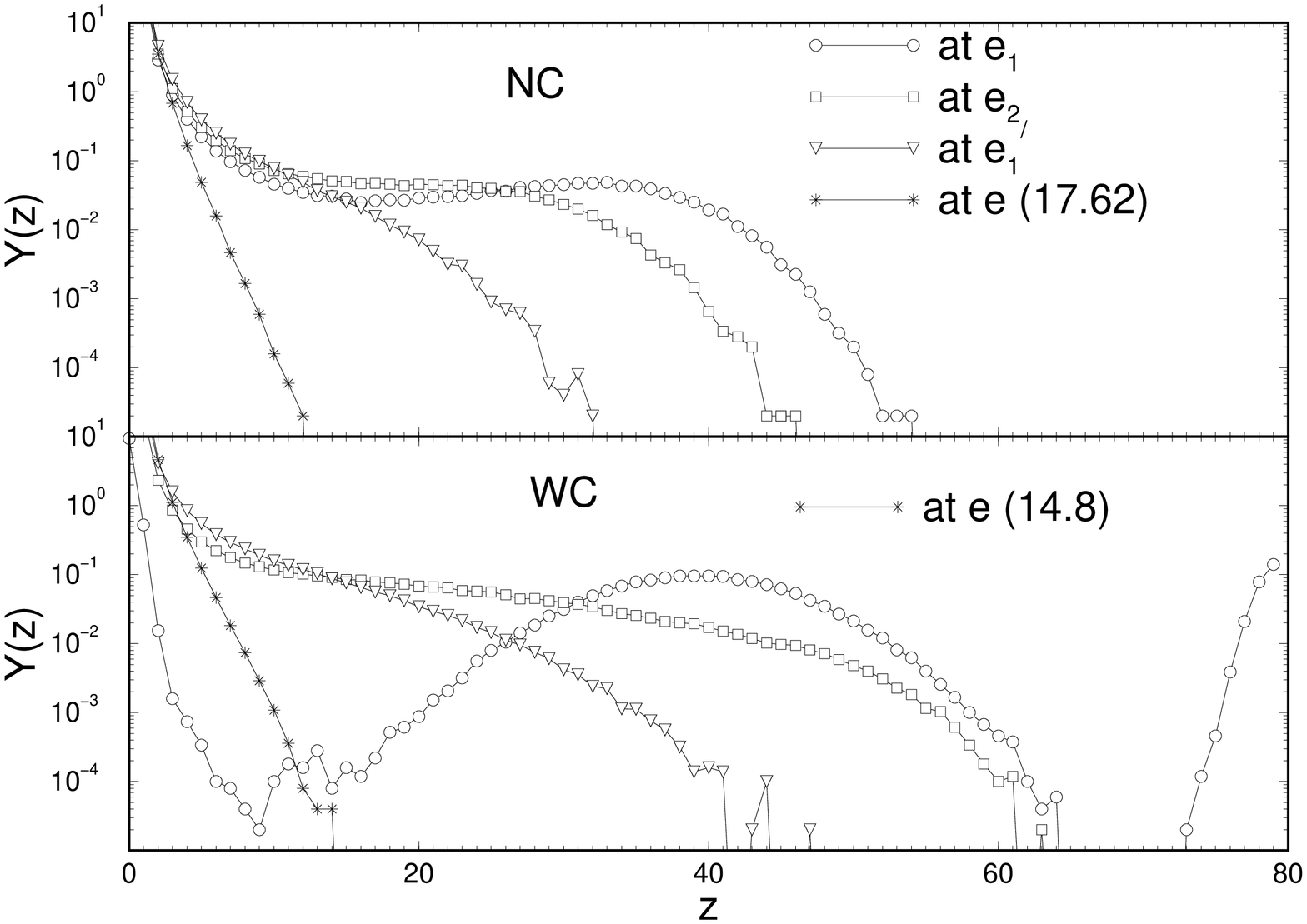}}}
\vskip 0.4in

Fig.2: {\it LGM simulations for 
$Y(z)$ {\it vs.} $z$ at $e_1, e_2$ and $e_1'$ without Coulomb
(top panel) and with Coulomb (bottom panel). In the top panel both
$e_1$ and $e_2$ are near the energy, where a maximum in the
yield at the high $z$ side has just disappeared. Qualitatively, this
marks the phase transition point. But
in the bottom panel where the Coulomb force is included, $e_1$ marks
an energy when there is still a large fragment. Thus this is below
the phase transition temperature. However $e_2$ still marks the
location when the maximum at the high $z$ side has just disappeared.
At much larger $e$ values (shown arbitrarily at $e=17.6$ MeV and $e=14.8$
MeV), $Y(z)$ falls much more rapidly with $z$.}

\vskip 0.2in
\epsfxsize=3.0in
\epsfysize=4.5in
\centerline{\rotatebox{270}{\epsffile{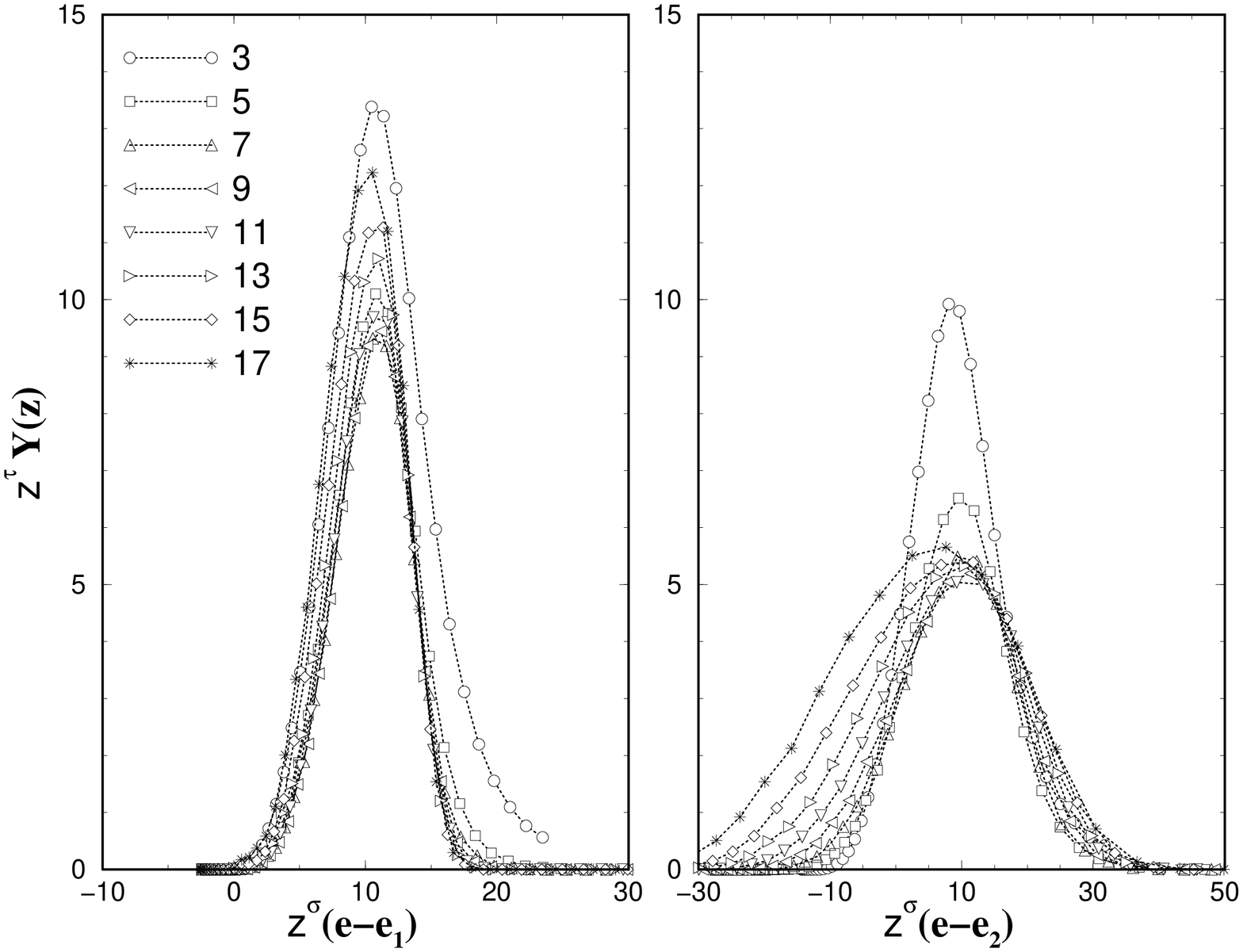}}}
\vskip 0.4 in

Fig.3: {\it For different isotopes $z$ we plot $z^{\tau}Y(z)$
against $z^{\sigma}(e-e_1)$ in the LGM, 
where the exponent $\tau$ is the ``best''
$\tau$ at $e_1$ and extraction of $\sigma, e_1$ is described in the text.
By scaling one means that curves for different $z$'s coalesce into
one.  This is approximately true for scaling around $e_1$ (left panel)
but not around $e_2$ (right panel).  Scaling around $e_1'$ is worse 
(not shown).}\\

In Fig. 3 we show that the scaling
law is rather well obeyed around $e_1$.  It is very poorly obeyed
around $e_1'$.  An interesting plot is a scaling law around
$e_2$.  This is also shown in Fig. 3.  Of course the scaling around
$e_2$ is nowhere as good as around $e_1$ but it is still better than
around $e_1'$ (not plotted).

\section{Calculations with a thermodynamic model}

Details of the thermodynamic model can be found in several places 
\cite{Dasgupta1,Dasgupta2,Bhatt}.  The physics assumption is
that composites are formed at an appropriate temperature at a  
volume larger than normal nuclear
volume dictated solely by consideration of phase space.  Thus the
model is close in spirit to the Statistical Multifragmentation Model
of Copenhagen \cite {Bondorf} with the simplification that the freeze-out
volume is assumed to be independent of the partitions.  This allows for
very quick computation without any Monte-Carlo simulations. 
The inputs for this calculation are the following.  Apart from neutrons and 
protons, experimental binding energies and ground state spins are
used for deuteron, triton, $^3$He and $^4$He.  No excited states
are included for these.  For mass 5 and higher we use the
semi-empirical formula for binding energies with volume term, surface
tension term, symmetry energy and Coulomb energy.  Excited states
for composites are included in the Fermi-gas approximation. The Coulomb
interaction between different composites is included in the 
Wigner-Seitz approximation \cite{Bondorf}.

\vskip 0.2in
\epsfxsize=3.5in
\epsfysize=5.0in
\centerline{\rotatebox{270}{\epsffile{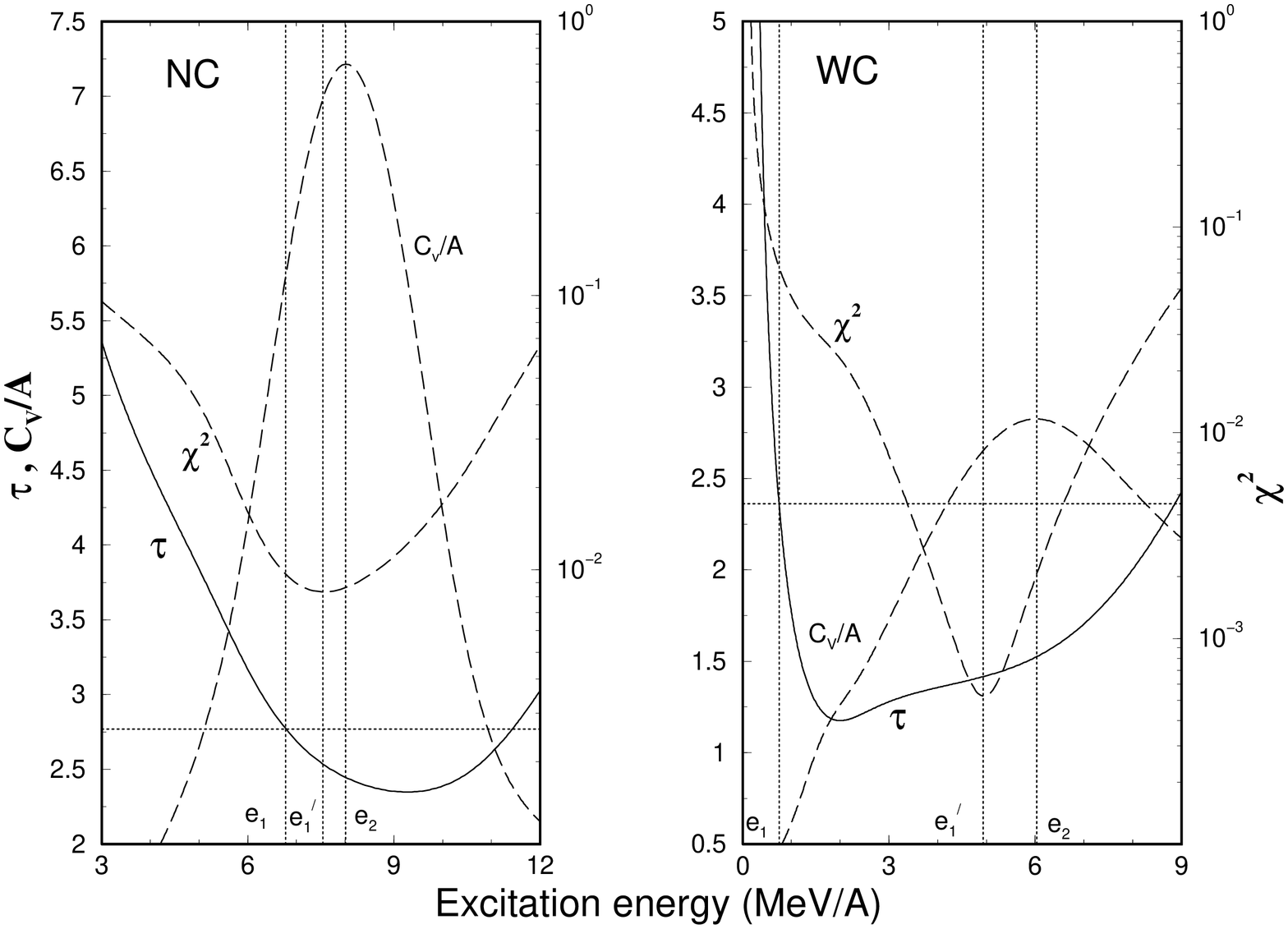}}}
\vskip 0.4 in

Fig.4: {\it Similar to Fig.1 except the calculation is with the thermodynamic
model (section III), with a freeze-out density $0.27\rho_0$.
Again note that with the inclusion of the Coulomb
force (WC), $\Delta e=e_2-e_1$ increases substantially compared to
no Coulomb (NC) case.}\\

Since this is a canonical calculation, calculations are done for
fixed temperatures.  For each temperature, the average excitation
energy per particle can be calculated.  For comparisons with
LGM, figures are drawn with energy as the abscissa. Table I
gives both the temperature and energy for relevant quantities.

There is no reason to expect results close to the ones calculated
using LGM.  For example, $e_1'$ is lower than $e_2$ in the thermodynamic
model but higher in the LGM.  Nonetheless, Fig. 4 shows that
$\Delta e=e_2-e_1$ increases significantly with inclusion of the
Coulomb force (from 1.24 MeV to 5.28 MeV).  Again $e_1$ does not
seem to mark the point of phase transition at all (Fig. 5) and
$e_2$ is a much better candidate.  This is so in spite of the
fact scaling is well obeyed with respect to $e_1$ and not so well
with respect to $e_2$ or $e_1'$ (Fig. 6). 

\vskip 0.2in
\epsfxsize=3.0in
\epsfysize=4.5in
\centerline{\rotatebox{270}{\epsffile{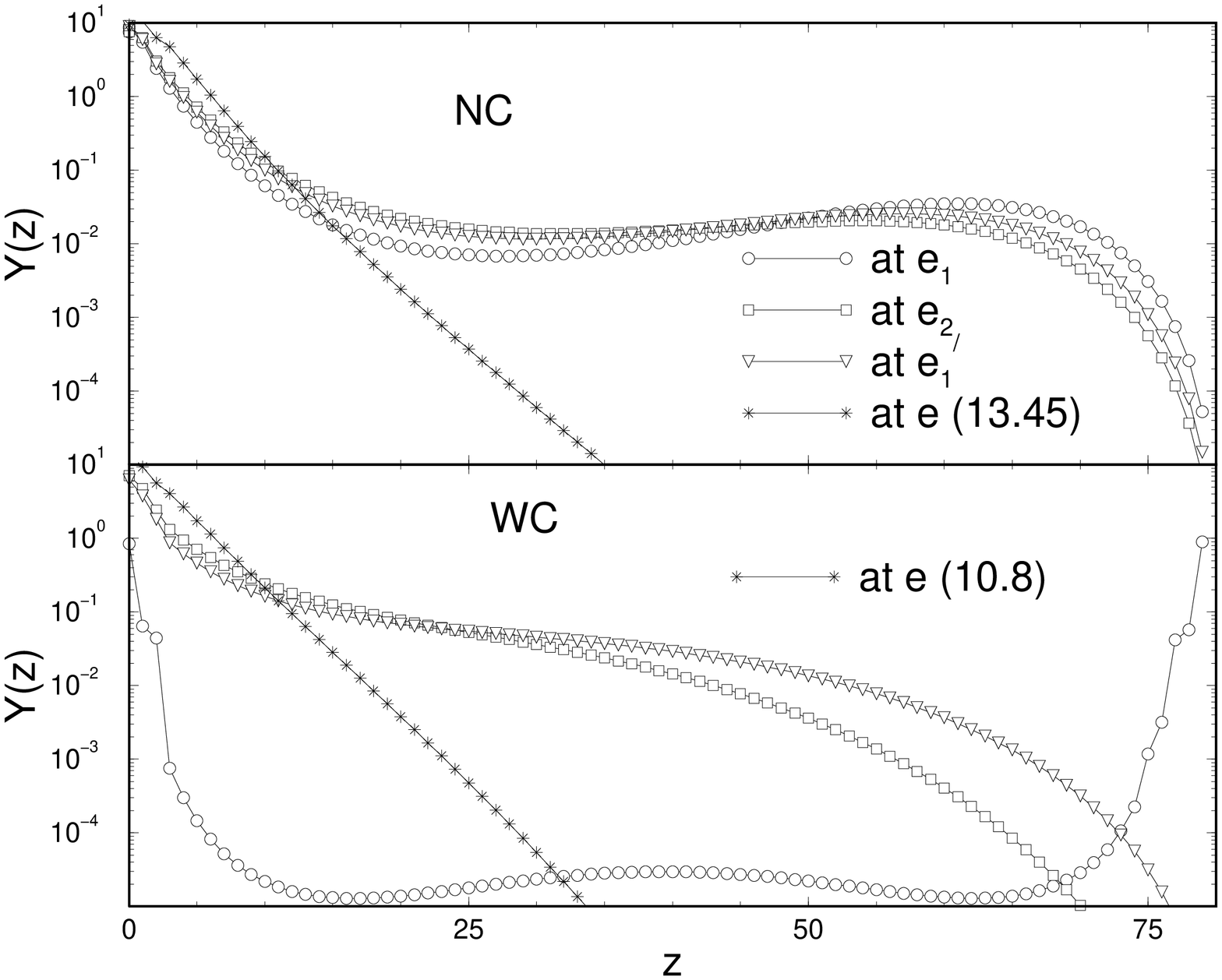}}}
\vskip 0.4 in

Fig.5: {\it Similar to Fig. 2 except the calculation is with the thermodynamic
model (section III).  Again with (WC) and without (NC) the Coulomb 
interaction, $e_2$ continues to be a
better mark for the point of phase transition energy.  Without the Coulomb both
$e_1$ and $e_2$ are acceptable.  At higher energies (chosen arbitrarily
at $e$=10.8 MeV and 13.45 MeV) the drops of yields with $z$ are much
faster.}

\epsfxsize=3.0in
\epsfysize=4.5in
\centerline{\rotatebox{270}{\epsffile{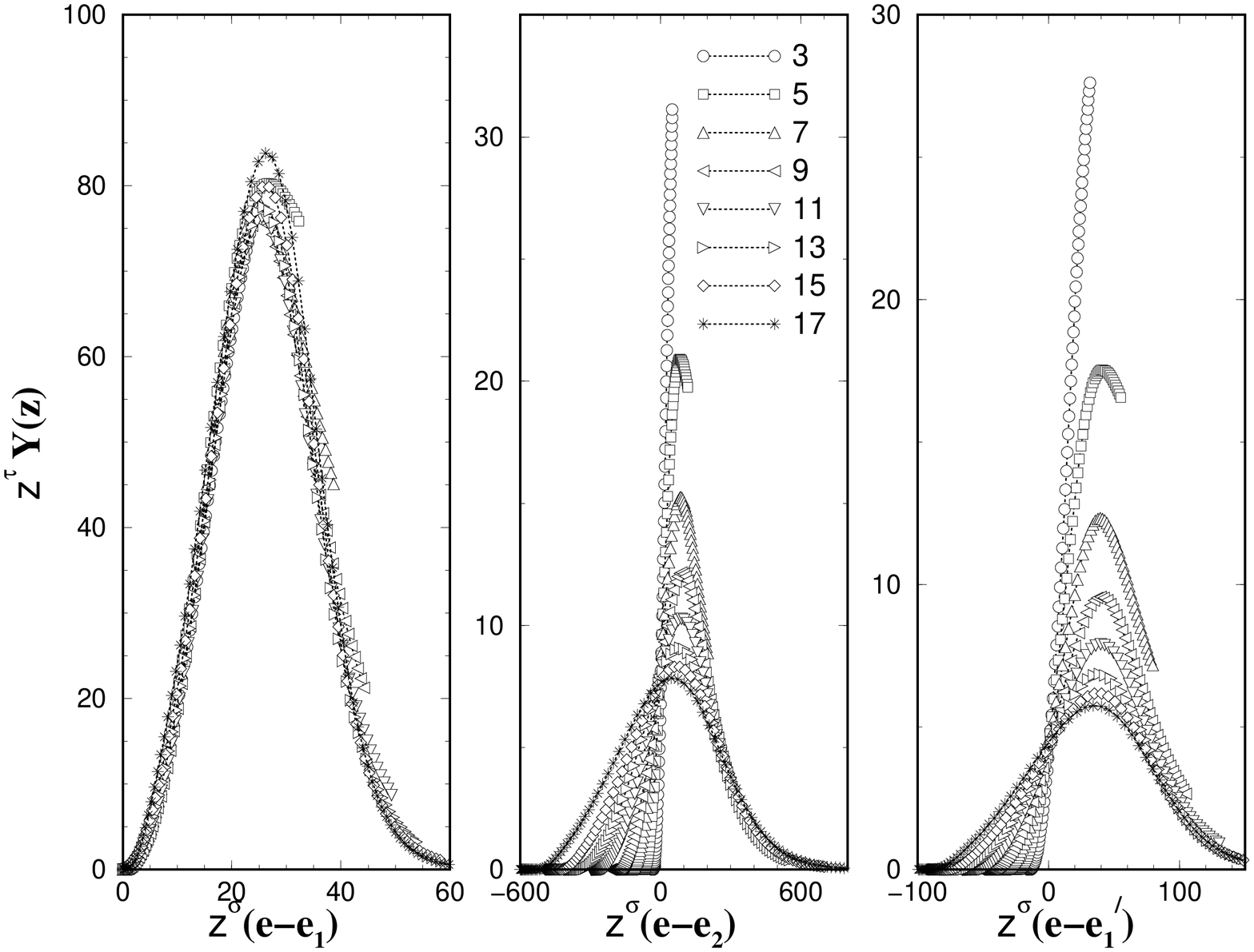}}}
\vskip 0.35 in

Fig.6: {\it Similar to Fig.3 except that the thermodynamic model is used
and scaling is tested around $e_1, e_2$ and $e_1$'.}

\section{8 GeV/c $\pi$$^-$ on Au data}

It is of interest to check if the conclusions reached in the
theoretical models are verified in experimental data 
\cite{Lefort,Beaulieu}.  Several non-trivial
issues need to be clarified before this can be attempted.  Because
the power-law fit is not exact, the extraction of $\tau$ 
from data or theoretical calculation has some ambiguity.  
In previous sections we calculated $\tau$
by minimising the quantity defined as $\chi^2$ in eq.(2.1) at each
$e$. But one could also minimise 
\begin{eqnarray}
\tilde{\chi}^2\equiv\sum (Y(z)-Cz^{-\tau})^2 .
\end{eqnarray}
If one is using experimental data, a more standard practice would be
to minimise \cite{Math}
\begin{eqnarray}
\hat{\chi}^2\equiv\sum\frac{(Y(z)-Cz^{-\tau})^2}{\sigma(z)^2} ,
\end{eqnarray}
where often the $\sigma$'s are statistical errors.  The difference in the
value of $\tau$ extracted by minimising Eq.(2.1) or Eq.(4.1) can 
be significant or small.  In theoretically calculated values of
$Y(z)$ the difference is small.  But it is not so in the experimental
values of $Y(z)$.
For the experimental data of 8 GeV/c $\pi^{-}$
on gold, the results of using Eqs. (2.1), (4.1) and (4.2)
are shown in Fig. 7.  One can give crude arguments that
minimising (2.1) rather than (4.1) means that $Y(z)$'s of higher
$z$'s are preferentially fitted.  In \cite{PLB}, Eq. (4.2) was
chosen.

In order to compare the experimental results to the theoretical calculations
we have taken the experimental yields $Y(z)$'s, ignored all errors
and repeated the calculations described in section II.  
We show in Fig. 7 the results of this analysis.
Qualitatively, the results are similar to the 
theoretical results (Figs.1 and 4). In those two figures the minimum 
in $\tau$ occurs at very low excitation energy for the WC case. 
For the experimental data the minimum, if it exists, is
also at a low value, below 1.5 MeV. More interesting is
the right panel of Fig. 7 where we plot the experimental specific
heat and find the maximum in specific heat coincides quite well
with the minimum of $\chi^2$ (this is $e_1'$).  This agrement is also quite
close to the theoretical predictions.  The specific heat was extracted by
differentiating with respect to $T$, the experimental caloric curve
obtained for the same data set by A. Ruangma et al. \cite{Ruangma}. 

\epsfxsize=3.5in
\epsfysize=5.0in
\centerline{\rotatebox{270}{\epsffile{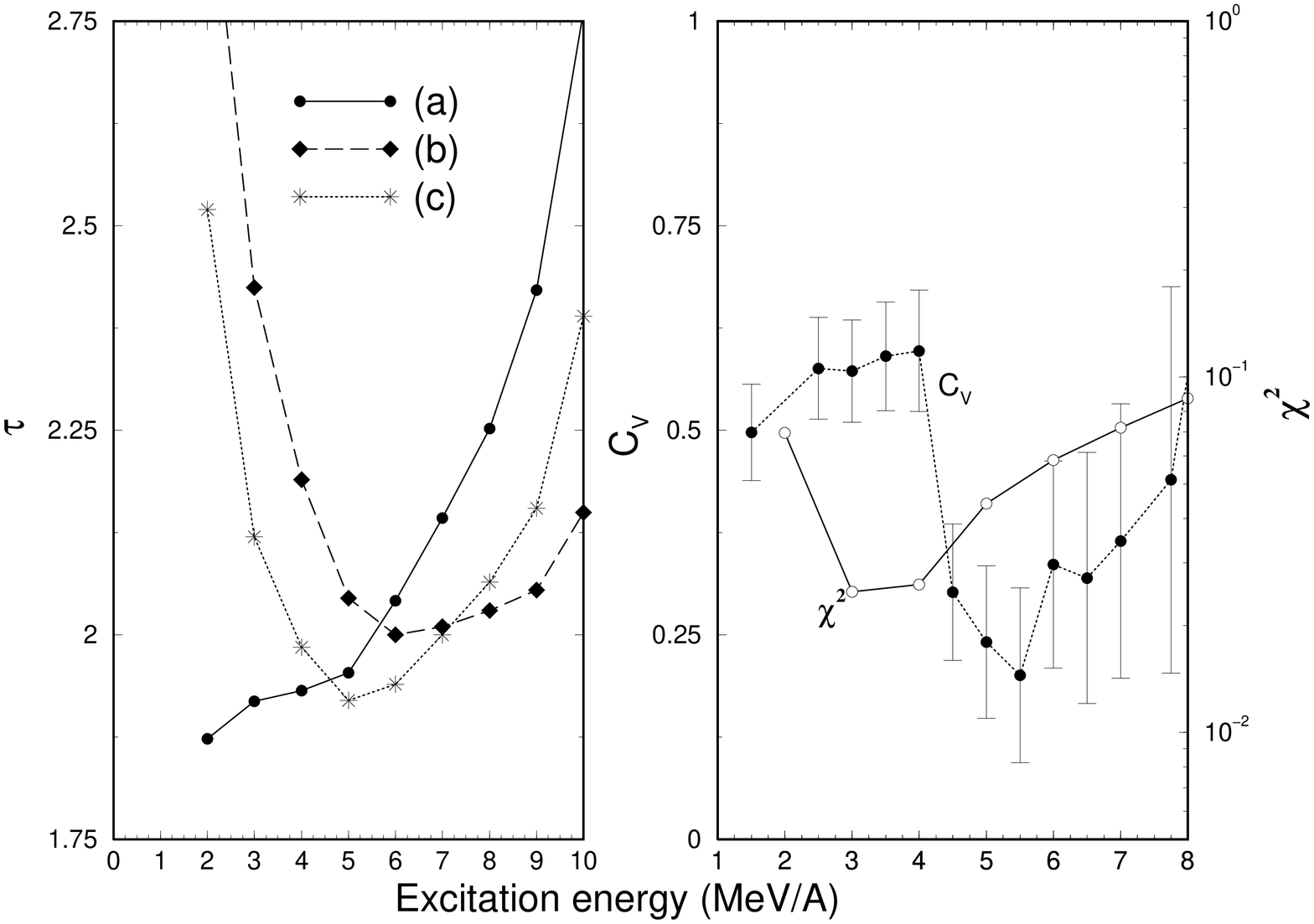}}}
\vskip 0.35 in

Fig.7: {\it The left panel shows the values of $\tau$ 
extracted from the ISiS data using
three different fomulae: (a) by minimising the right hand side of
Eq.(2.1) at each $e$ value, (b) by minimising the right hand side of 
Eq.(4.1) and (c) by minimising the right hand side of Eq.(4.2).  In
the right panel $C_V$ from data and $\chi^2$ (Eq.(2.1)) calculated
from data are plotted. The minimum of $\chi^2$ and maximum of $C_V$ 
coincide within experimental uncertainty.}\\

We also test the scaling around $e_1'$ (Fig. 8)
although some qualifying comments need to be made about this figure.
In experiments, the source size as well as the charge of the thermal
source, depend upon $e$.  The scaling law, which spans $e$ values on
either side of $e_1'$, assumes constant source size as well constant
charge.  Thus the scaling law can not be directly tested without
additional corrections renormalising the yields $Y(z)$ to 
compensate for changes in the source size and charge. 
This was not done here. (However the $\tau$ values and
the values of $\chi^2$ should be insensitive to such changes although
we do require that for a given $e$, the source size and charge remain
unchanged.  This last condition is approximately obeyed.)

We can summarise the results of the comparison with experimental data 
as follows.
In the data the maximum of specific heat ($e_2$) and the minimum of
$\chi^2$, $e_1'$ both are $e\approx 4 MeV$
with a value of $\tau$ about 2.1 and $\sigma$ about 0.53 . 
This is very close to to the results of Elliott et al. \cite{Elliott}
for the same data set using Fisher's droplet model approach and 
Berkenbusch et al. 
\cite{Berkenbusch} using a percolation model.  
For percolation, the excitation energy was defined as the critical
excitation energy in the sense of a second order phase transition.
The thermodynamic model, which has no adjustable parameters and only
a first order phase transition, reproduces trends of the data very well
although both $e_2$ and $\sigma$ are higher, 6 MeV and 0.96 respectively.

\epsfxsize=3.0in
\epsfysize=3.5in
\centerline{\rotatebox{270}{\epsffile{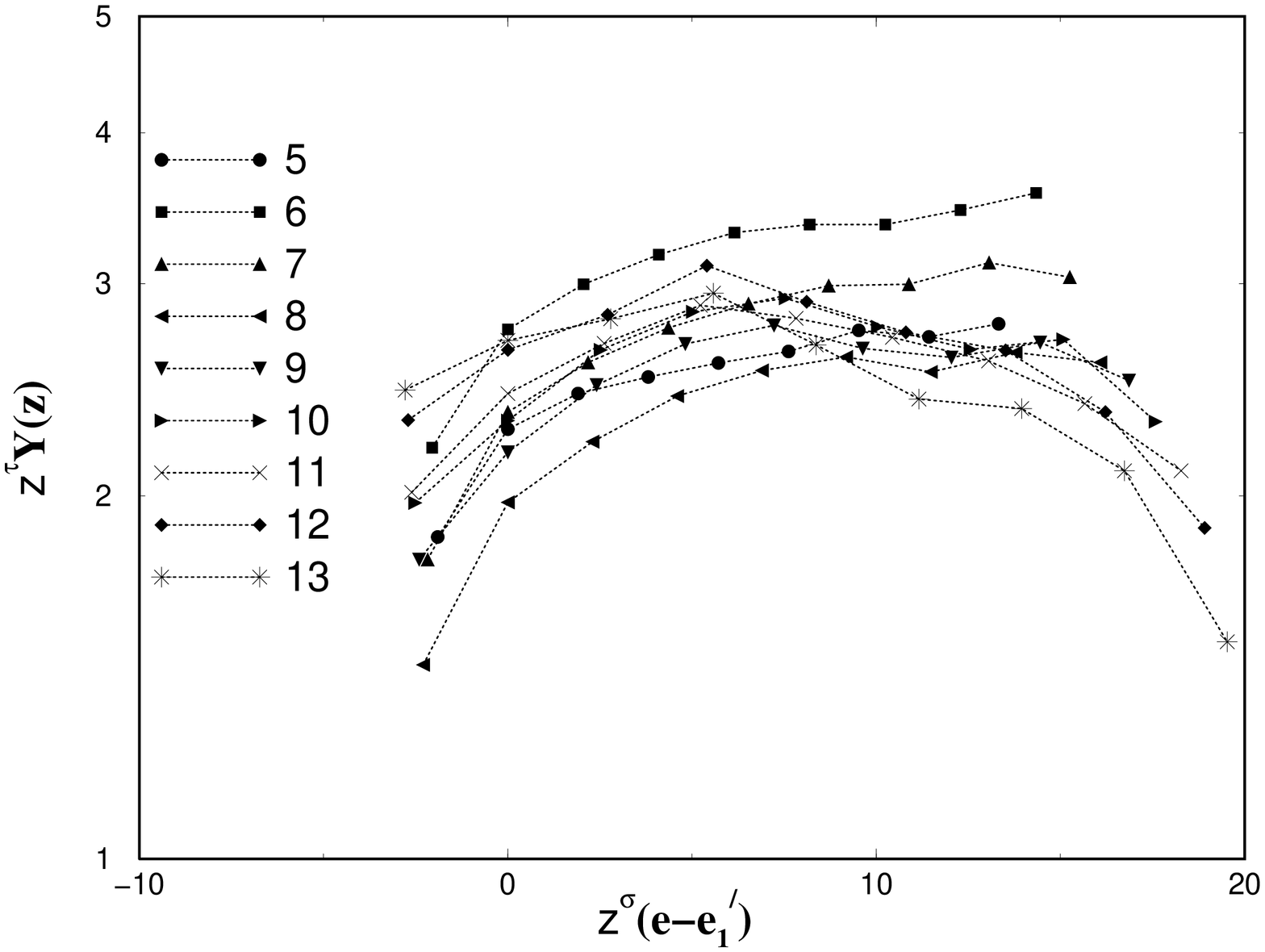}}}
\vskip 0.2 in

Fig.8: {\it Scaling behavior associated with the ISiS data, 
around $e_1'$.}\\

Since source sizes change with excitation per particle, we tested the
sensitivity of model calculations with regard to size using the
thermodynamic model.  Table II gives the results. Within the range of 
variation of relevance to the experimental data the changes are small,
though not negligible.

\section{Summary and Conclusion}

We have discussed three characteristic excitation energies (equivalently
temperatures): two of them have their origin in Eq.(1.2) ($e_1$=
value of $e_0$ which gives best scaling and $e_1'$=value of $e_0$ where
deviation from power-law is the least)
and the third, $e_2$,
is obtained from an extremum in specific heat.  In an ideal situation,
all three would have the same value.  In theoretical models
they are close if the Coulomb force is omitted.  The Coulomb force
makes a substantial splitting between $e_1$ and $e_1'$, indicating
the sensitivity of the extracted thermodynamic quantities to this 
interaction.  We find
that $e_1'$ gives a better measure of the phase transition energy
and it stays close to $e_2$.  In the experimental data that we 
considered $e_1'$ and $e_2$ follow this pattern.

The quantitative results are dependent upon freeze-out densities
and the source sizes but not sensitively so.

Lastly, we have not discussed the order of phase transition but the 
model calculations in sections 2 and 3  imply a first order phase 
transition.  In this interpretation, depending upon the 
excitation energy, most fragments are emitted while inside the 
co-existence region of the phase diagram (and possibly the spinodal
region) and the extracted ``critical'' excitation energy indicates
the boiling point.  As pointed out earlier, the boiling point is at a
similar excitation energy as the critical point found by Elliott et al.
\cite{Elliott} from an analysis based on the Fisher droplet model and
Berkenbusch et al. \cite{Berkenbusch} based on percolation theory.
Further discussions of critical phenomena \cite{Elliott,Berkenbusch} 
in disassembly of hot nuclei as opposed to first order phase transition
in the disassembly can be found in \cite{Das2,pan,gul}

A first order phase transition is consistent with recent observations
by the ISiS collaboration \cite{Beaulieu,beau,lef} of a strong increase in
fragment production probability, a strong decrease in fragment emission
time, and the onset of collective radial expansion above 4A MeV of excitation
energy, which were interpreted as signatures for bulk emission.  It
is also consistent with the flattening of the caloric curve from
which the heat capacity was extracted \cite{Ruangma}.

\section{Acknowledgment}
This work is supported in part by the Natural Sciences and Engineering
Research Council of Canada and by {\it le Fonds pour la Formation
de Chercheurs et l'aide \`a la Recherche du Qu\'ebec}. Experiment E900
was supported by the US Department of Energy and the National Science
Foundation.

\begin{table}
\caption{The values of the parameters $e_1 , e_2$ and $e_1'$, corressponding
$T_1 , T_2$ and $T_1'$; and  $\tau$'s at these $e_1 , e_2 , e_1'$ at
$e_1$ as obtained in LGM and thermodynamic models, for freeze-out
density of $0.27\rho_0$.
Values are shown for calculations with (WC) and without (NC) Coulomb
interactions.}
\vspace {0.5in}
\begin{tabular}{ccccc}
\hline
\multicolumn{1}{c}{Parameters} &
\multicolumn{1}{c}{LGM (NC)} &
\multicolumn{1}{c}{LGM (WC)} &
\multicolumn{1}{c}{THDM (NC)} &
\multicolumn{1}{c}{THDM (WC)}\\
\hline
$e_1$&11.66&3.60&6.77&0.75\\
$T_1$&4.46&1.95&7.53&2.94\\
$\tau (e_1)$&1.964&1.771&2.77&2.36\\
$e_2$&12.61&8.38&8.01&6.03\\
$T_2$&4.62&3.38&7.715&6.445\\
$\tau (e_2)$&1.824&1.251&2.44&1.55\\
$e_1'$&13.88&10.5&7.54&4.94\\
$T_1'$&4.88&3.82&7.65&6.05\\
$\tau (e_1')$&2.544&1.851&2.53&1.42\\
\hline
\end{tabular}
\end{table}

\begin{table}
\caption{The values of the parameters for two different sizes
of the fragmenting source as
obtained in the thermodynamic model, for a freeze-out density of
$0.27\rho_0$.}
\vspace {0.5in}
\begin{tabular}{ccc}
\hline
\multicolumn{1}{c}{Parameters} &
\multicolumn{1}{c}{N=118} &
\multicolumn{1}{c}{N=101} \\
\multicolumn{1}{c}{} &
\multicolumn{1}{c}{Z=79} &
\multicolumn{1}{c}{Z=68} \\
\hline
$e_1$&0.75&1.01 \\
$T_1$&2.94&3.46 \\
$\tau (e_1)$&2.36&2.35 \\
$e_2$&6.03&5.90 \\
$T_2$&6.445&6.40 \\
$\tau (e_2)$&1.55&1.49 \\
$e_1'$&4.94&4.91 \\
$T_1'$&6.05&6.05 \\
$\tau (e_1')$&1.42&1.41 \\
\hline
\end{tabular}
\end{table}


\begin{references}
\bibitem{Dasgupta1} S. Das Gupta, A. Z. Mekjian and M. B. Tsang, Advances
in Nuclear Physics, vol.26, 91 (2001).

\bibitem{Stauffer} D. Stauffer and A. Aharony, {\it Introduction to
Percolation Theory} (Taylor and Francis, London, 1992).

\bibitem{Poch} J. Pochodzalla et al., Phys. Rev. Lett. {\bf 75}, 1040
(1995).

\bibitem{Co} A Coniglio and W. Klein, J. of Phys. {\bf A13}, 2775 (1980)
223(1981).

\bibitem{Pan} J. Pan and S. Das Gupta, Phys. Rev {\bf C51},1384 (1995).

\bibitem{Dasgupta2} S. Das Gupta and A. Z. Mekjian, Phys. Rev {\bf C57},
1361 (1998).

\bibitem{Bhatt} P. Bhattacharyya, S. Das Gupta and A. Z. Mekjian,
Phys. Rev {\bf C60}, 54616 (1999).

\bibitem{Bondorf} J. P. Bondorf, A. S. Botvina, A. S. Iljinov, I. N.
Mishustin, and K. Sneppen, Phys. Rep. {\bf 257}, 133 (1995).

\bibitem{Das1} C. B. Das, S. Das Gupta and S. K. Samaddar, Phys. Rev. 
{\bf C63}, 011602(R), (2001).

\bibitem{Sam} S. K. Samaddar and S. Das Gupta, Phys. Rev {\bf C61}, 034610
(2000).

\bibitem{Campi} X. Campi and H. Krivine, Nucl. Phys. {\bf A620}, 46 (1997).


\bibitem{Das2} C. B. Das, S. Das Gupta and A. Majumder, Phys. Rev. {\bf C65},
034608 (2002).

\bibitem{Gulminelli} F. Gulminelli and Ph. Chomaz, Phys. Rev. Lett. 
{\bf 82}, 1402 (1999).

\bibitem{Scharenberg} R. P. Scharenberg et al., Phys. Rev. {\bf C64}, 054602
(2001).

\bibitem{Lefort} T. Lefort et al., Phys. Rev {\bf C64}, 064603 (2001).

\bibitem{Beaulieu} L. Beaulieu et al., Phys. Rev. {\bf C64}, 064604 (2001).

\bibitem{Math} W. H. Press, B. P. Flannery, S. A. Teukolsky and W. T.
Vetterling, {\it Numerical Recipes}, (Cambridge University Press,
New York, 1987) pp. 504 .

\bibitem{PLB} L. Beaulieu et al., Phys. Lett. {\bf B463}, 159 (1999).

\bibitem{Ruangma} A. Ruangma et al., nucl-ex/010004, submitted to PRC.

\bibitem{Elliott} J. B. Elliott et al., Phys Rev. Lett {\bf 88},
042701 (2002)

\bibitem{Berkenbusch} M. K. Berkenbusch et al., Phys. Rev. Lett.
{\bf 88}, 022701 (2002)

\bibitem{pan} J. Pan, S. Das Gupta and M. Grant, Phys. Rev. Lett.
{\bf 80} 1182 (1998)

\bibitem{gul} F. Gulminelli, Ph. Chomaz, M. Bruno and M. D'Agostino,
Phys. Rev.{\bf C65}, R 051601 (2002)

\bibitem{beau} L. Beaulieu et al., Phys. Rev. Lett {\bf 84}, 5971 (2000)

\bibitem{lef} T. Lefort et al., Phys. Rev {\bf C62}, R031604 (2000)
\end{references}
\end{document}